\RequirePackage{fix-cm}
\documentclass[twocolumn]{svjour3}          
\smartqed  
\usepackage{graphicx}
\usepackage{amsmath}
\usepackage{bm}
\usepackage{amsfonts}
\usepackage{amssymb}
\usepackage{mathtools}
\usepackage{multirow}
\usepackage{url}
\usepackage{algorithm} 
\usepackage{algpseudocode}
\usepackage[draft]{fixme}
	
\newcommand{\textsub}[2]{{#1}_{\text{#2}}}
\newcommand{\textsup}[2]{{#1}^{\text{#2}}}
\makeatother

\newcommand{\T}{{\text{T}}}
\newcommand{\tv}[1]{\textbf{#1}}

\newcommand{\tm}[1]{\textbf{#1}}

\newcommand{\tran}[1]{^{\text{#1}}}
\newcommand{\bg}[1]{\boldsymbol#1}

\journalname{Computational Particle Mechanics}
\begin{document}

\title{Warm starting the projected Gauss-Seidel algorithm for granular matter simulation}

\titlerunning{Warm starting the projected Gauss-Seidel algorithm}        

\author{Da Wang \and Martin Servin \and Tomas Berglund}


\institute{D.~Wang \at
              Ume\aa\ University
           \and
           M.~Servin \at
              Ume\aa\ University \\
              Tel.: +46-90-7866508\\
              \email{martin.servin@umu.se}           
           \and
           T.~Berglund \at
              Algoryx Simulation AB
}

\date{Received: date / Accepted: date}

\maketitle

\begin{abstract}
The effect on the convergence of warm starting the projected Gauss-Seidel solver
for nonsmooth discrete element simulation of granular matter are investigated.
It is found that the computational performance can be increased by a factor
2 to 5. 
\keywords{Discrete elements \and Nonsmooth contact dynamics \and Convergence \and Warm starting \and Projected Gauss-Seidel}
\end{abstract}

\section{Introduction}
\label{sec:introduction}

In simulations of granular matter using the nonsmooth discrete element method (NDEM) \cite{Moreau:1999:NAS,Jean:1999:NSC,Radjai:2009:cdn} the computational time is dominated by the solve stage, where the contact forces
and velocity updates are computed.  Conventionally this involves solving a mixed complementarity problem
or a quasi-optimization problem that arises from implicit integration of the rigid multibody equations of motion in
conjunction with set-valued contact laws and impulse laws, usually the Signorini-Coulomb law and Newton impulse
law.  The computational properties of the solution algorithms for these problems are largely open questions,
lacking general proof of existence and uniqueness of solutions as well as of general proof of convergence and 
numerical stability \cite{brogliato:2002:NSF}.  The projected Gauss-Seidel (PGS) algorithm is widely used.
The popularity of PGS is likely due to having low computational cost per iteration, small memory footprint 
and produce smooth distribution of errors that favour stable simulation.  
In many cases PGS require few iterations to identify the active set of constraints.
This make PGS a natural choice for fast simulations of large-scale rigid multibody systems with frictional
contacts.  The asymptotic convergence, however, is slow.  The PGS algorithm solves each local two-body contact problem accurately but approaches
to the global solution in a diffusive manner with iterations.  This limit
the practical use of PGS for simulations of high accuracy.   
The residual error appear as artificial elasticity \cite{unger:ebc:2002},
with an effective sound velocity $\textsub{v}{PGS} = \sqrt{\textsub{N}{it}}d/\Delta t$, where $\textsub{N}{it}$ is the number of iterations, $d$ is the
particle size and $\Delta t$ is the timestep.  Accurate resolution of the 
impulse propagation in stiff materials thus require large number of iterations
or small timestep.  The required number of iterations for a given error tolerance
increase with the size of the contact network, particularly with the number of contacts in 
direction of gravity or applied stress.  It may, however, 
saturate by an arching phenomena analogous to the Janssen's law for silos \cite{servin:2014:esn}.
The PGS algorithm is parallelizable, despite many many authors claim of the opposite,
for hardware with distributed memory using domain decomposition methods
\cite{precklik:usn:2015,visseq:hpc:2013}.

\begin{figure}
  \includegraphics[width=0.45\textwidth]{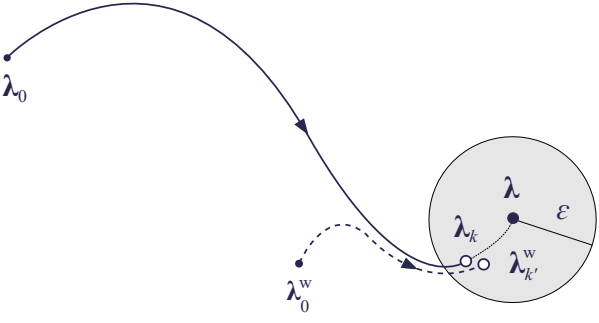}
\caption{Illustration of improved convergence by warm starting.}
\label{fig:lambda_converge}
\end{figure}

Warm starting means to start the PGS algorithm with an initial guess, 
$\bm{\lambda}_0^{\text{w}}$, that presumably is closer to 
the exact solution, $\bm{\lambda}$ , than starting with the nominal choice of 
$\bm{\lambda}_0 = 0$.  The idea, illustrated in Fig.~\ref{fig:lambda_converge}, is that the warm started PGS
reach an approximate solution, $\bm{\lambda}_{k'}$, with fewer
iterations than the solution, $\bm{\lambda}_{k}$, starting 
from nominal value. In other words,  
$|\bm{\lambda} - \bm{\lambda}_{k'}^{\text{w}}| \lesssim |\bm{\lambda} - \bm{\lambda}_{k}| < \varepsilon$ 
with $k'  <  k$ and error tolerance $\varepsilon$.  
The effective increase in convergence should be most significant for static or nearly static
configurations.  For rapid granular flows the solution change rapidly with time and no or little effect on
convergence is expected.  There have been several reports on improved convergence by using warm starting 
\cite{Radjai:2009:cdn,kaufman:2008:spf,kaufman:2009:cpc,erleben:2013:nml,moravanszky:2004:fcr,todorov:2010:inc}
but to the best of our knowledge no quantitative analysis has previously been presented.

\section{PGS for nonsmooth discrete element simulation}
\label{sec:PGS}

The mixed complementarity problem (MCP) for computing the update of the velocity
from $\textsub{\tv{v}}{old}  \equiv \bm{v}(t - \Delta t) $ to $\tv{v}  \equiv \bm{v}(t)$
and the Lagrange  multiplier $\bm{\lambda}$ of the contact constraints and take the form
\begin{equation}
\label{eq:MLCP}
	\left[ 
	    \begin{array}
		    [c]{cc}%
		    \tm{M} & - \tm{G}^\T \\
		    \tm{G} & \bg{\Sigma}
	    \end{array}    
	\right]
	\left[ 
	    \begin{array}
		    [c]{c}%
		    \tv{v} \\
		    \bg{\lambda}
	    \end{array}    
	\right]
	=
	\left[ 
	    \begin{array}
		    [c]{c}%
		    \tv{p}\\
		    \tv{q}
	    \end{array}    
	\right]	
\end{equation}
\begin{equation}
	\bg{\lambda}^{(\alpha)}\in \mathcal{C}_{\mu}( \textsub{\lambda}{n}^{(\alpha)})\ \ ,\ \alpha = 1,2,\hdots,\textsub{N}{c}
\end{equation}
where  $\bm{M}$ is the mass matrix and $\bm{G}$ the Jacobian of 
contact constraints. The contact force, $\bm{G}^T \bm{\lambda}$,
is restricted by a friction cone condition that we represent 
$\bg{\lambda}^{(\alpha)}\in  \mathcal{C}_{\mu}(  \textsub{\lambda}{n}^{(\alpha)})$,
where $\alpha$ indexes the contacts.
The diagonal perturbation $\bm{\Sigma}$ regularize the problem and allow modeling of contact elasticity.  The vectors 
$\bm{p}$ and $\bm{q}$ on the right hand side depend on particle inertia, external force and 
constraint violations on position and velocity level.  As friction cone condition we use the
Signorini-Coulomb law including $0 \leq \textsub{\lambda}{n}^{(\alpha)}$
and $|\bg{\lambda}_{\text{t}}^{(\alpha)}| \leq \textsub{\mu}{s}|\tm{G}_{\text{n}}^{(\alpha)\T} \lambda_{\text{n}}^{(\alpha)}|$
with the friction coefficient $\textsub{\mu}{s}$ for each contact
$\alpha$ divided in one normal (n) and two tangential (t) components.  The constraint forces act to
prevent contact overlap, $\bg{g}\leq 0$, and contact sliding, $\textsub{\bg{G}}{t} \bg{v} = 0$.
Similarly, rolling resistance (r) is imposed by a constraint $\textsub{\bg{G}}{r} \bg{v} = 0$
with a Coulomb like law: $|\bg{\lambda}_{\text{r}}^{(\alpha)}| \leq \textsub{\mu}{r}r^{*}|\tm{G}_{\text{n}}^{(\alpha)\T} \lambda_{\text{n}}^{(\alpha)}|$, where $r^*$ is the effective radius.  
See Appendix A for further details.  For a system with $\textsub{N}{p}$ 
particles represented as rigid bodies and $\textsub{N}{c}$ contacts with
normal and tangential force and rolling and twisting resistance the vectors and matrices in
Eq.~(\ref{eq:MLCP}) have the following dimensions $\dim (\bm{M}) = 6\textsub{N}{p}\times6\textsub{N}{p}$,
$\dim (\bm{G}) = 6\textsub{N}{c}\times\textsub{N}{p}$, $\dim (\bm{}v) = \dim (\bm{p}) = 6\textsub{N}{p}$
$\dim (\bm{\lambda}) = \dim (\bm{q}) = 6\textsub{N}{c}$.  The matrices are however very sparse.
$\bm{M}$ and $\bm{\Sigma}$ are block diagonal and $\bm{G}$ is block sparse.  The blocks have dimension $6 \times 6$.
The main steps of the PGS iteration are
\begin{eqnarray}
	\bg{\lambda}^{(\alpha)}_{k+1} & = & \bg{\lambda}^{(\alpha)}_{k} + D^{-1}_{(\alpha\alpha)} \tv{r}^{(\alpha)}_{k}\\
	\bg{\lambda}^{(\alpha)}_{k+1} & \leftarrow & \text{proj}_{\mathcal{C}_{\mu}}(\bg{\lambda}^{(\alpha)}_{k+1}) \\
	\tv{v}_{k+1} & = & \tv{v}_{k} + \tm{M}^{-1} \tm{G}_{(\alpha)}^\T\Delta\bg{\lambda}_{k+1}^{(\alpha)}
\end{eqnarray}
with iteration index $k = 0,1,2,\hdots,\textsub{N}{it} - 1$,
change in multiplier $\Delta\bg{\lambda}_{k+1}^{(\alpha)} = \bg{\lambda}_{k+1}^{(\alpha)} - \bg{\lambda}_{k}^{(\alpha)}$ and residual
\begin{equation}\label{eq:residual}
	\tv{r}^{(\alpha)}_k = \tm{S}_{(\alpha\alpha)}\bg{\lambda}^{(\alpha)}_k 
	+ \tm{G}_{(\alpha)}\tm{M}^{-1}\tv{p}_{(\alpha)}-\tv{q}_{(\alpha)} = \tm{G}_{(\alpha)}\tv{v}_k - \tv{q}_{(\alpha)}
\end{equation}
where $\tv{v}_k \equiv \tm{M}^{-1}\tv{p} + \tm{M}^{-1}\tm{G}_{(\alpha)}^\T\bg{\lambda}_{k}^{(\alpha)}$
and $\tm{D}$ is the block diagonal part of the Schur complement matrix $\tm{S} = \tm{G}\tm{M}^{-1}\tm{G}^\T + \bg{\Sigma}$.
The details of the vectors $\tv{p}$ and $\tv{q}$ depend on the stepping scheme and constraint stabilization method.
When integrating with fix timesteps $\Delta t$ using the SPOOK stepper \cite{lacoursiere:2007:rvs} one has 
$\tv{p} = \tm{M}\textsub{\tv{v}}{old} + \Delta t \textsub{\tv{f}}{ext}$, with
smooth external forces $\textsub{\tv{f}}{ext}$,
and $\tv{q} = (\tv{q}^T_{\text{n}} , \tv{q}^T_{\text{t}}, \tv{q}^T_{\text{r}})^T$ with $\tv{q}_{\text{n}} = -(4/\Delta t)\bg{\Upsilon}\bar{\tv{g}} + \Upsilon\tm{G}_{\text{n}}\textsub{\tv{v}}{old}$, 
$\tv{q}_{\text{t}} = \tv{0}$ and $\tv{q}_{\text{r}} = \tv{0}$.  The projection $\bg{\lambda}^{(\alpha)}_{k+1} \leftarrow \text{proj}_{\mathcal{C}_{\mu}}(\bg{\lambda}^{(\alpha)}_{k+1})$ is made by simply clamping
$\bg{\lambda}^{(\alpha)}_{k+1}$ to the friction or rolling resistance limit if exceeded.
After stepping the velocities and positions an impact stage follows.  This include solving a 
MCP similar to Eq.~(\ref{eq:MLCP}) but with the Newton impact law,
$\textsub{\tm{G}}{n}^{(\alpha)} \tv{v}_{+} = - e\textsub{\tm{G}}{n}^{(\alpha)} \tv{v}_{-}$,
replacing the normal constraints for the contacts with normal
velocity larger than an impact velocity threshold $\textsub{v}{imp}$.
The remaining constraints are maintained by imposing $\tm{G}\tv{v}_{+} = 0$.
An algorithm of NDEM simulation with PGS is given in Appendix A together with details on
the Jacobians and relation between the solver parameters and material parameters.

\section{PGS warm starting}
\label{sec:warmstarting}
By default the PGS algorithm is initialized with $\bg{\lambda}^{(\alpha)}_{0} = \tv{0}$.
We refer to this as \emph{cold starting}.
In a stationary state the contact force $\tm{G}^T \bg{\lambda}$ is constant in time. 
In a nearly stationary state we expect the multipliers to remain almost constant
between two timestep.  Therefore it is reasonable to use the solution from last
timestep as an initial guess, $\bg{\lambda}(t) \approx \bg{\lambda}(t - \Delta t)$.
We use a fraction $\beta = 0.85$ of the solution from last timestep
\begin{equation}
\label{eq:history_warmstart}
	\bg{\lambda}_{0}(t) = \beta \bg{\lambda}_{\textsub{N}{it}}(t - \Delta t)
\end{equation}
It is important to also apply the corresponding impulse to the particles
and update the velocity
\begin{equation}
\tv{v}_0 = \tm{M}^{-1}\tv{p} + \tm{M}^{-1}\tm{G}^\T\bg{\lambda}_{0}
\end{equation}
such it become consistent with the initial guess for the multiplier.  We refer to warm starting
based on the last solution as \emph{history based warm starting}.  For any new contact we set 
$\bg{\lambda}^{(\alpha)}_{0} = \tv{0}$.  Warm starting is not applied at the impact stage
and we assume that the contact network is not fundamentally rearranged by the impacts
and use the solution from last timestep despite the occurrence of impacts.  

An alternative method for warm starting a nearly stationary state is to estimate each 
local contact force and assign this to the contact multipliers.  When using regularized NDEM 
the local contact force can be estimated from the overlaps and relative contact velocities
much as in conventional smooth DEM.  We refer to this approach as \emph{model based warm starting}.
For normal forces we use the Hertz contact law $\textsub{f}{n} = \textsub{k}{n}g^{3/2}_{\text{n}}$,
with overlap function $g_{\text{n}}$ and based on $\textsub{\tv{f}}{n} \approx \textsub{\tm{G}}{n}^T \textsub{\bg{\lambda}}{n}/\Delta t$ we estimate  
\begin{equation}
	\lambda_{\text{n},0}^{(\alpha)} \approx \tfrac{5}{4}\Delta t \textsub{k}{n}g^{5/4}_{\text{n}(\alpha)} 
\end{equation}
Similarly the regularized tangent friction force and rolling resistance force can be estimated via the Rayleigh dissipation functions
to
\begin{eqnarray}
	\bg{\lambda}_{\text{t},0} & \approx & \textsub{\gamma}{t}^{-1}\Delta t(\textsub{\tm{G}}{t}\tv{v})^\T\textsub{\tm{G}}{t}\\
	\bg{\lambda}_{\text{r},0} & \approx & \textsub{\gamma}{r}^{-1}\Delta t(\textsub{\tm{G}}{r}\tv{v})^\T\textsub{\tm{G}}{r}
\end{eqnarray}
Note that the friction and rolling resistance should, if large, be clamped to obey the conditions
$|\bg{\lambda}_{\text{t}}^{(\alpha)}| \leq \textsub{\mu}{s}|\tm{G}_{\text{n}}^{(\alpha)\T} \lambda_{\text{n}}^{(\alpha)}|$
and $|\bg{\lambda}_{\text{r}}^{(\alpha)}| \leq \textsub{\mu}{r}r^{*}|\tm{G}_{\text{n}}^{(\alpha)\T} \lambda_{\text{n}}^{(\alpha)}|$.

\section{Numerical experiments}
\label{sec:experiments}
Numerical simulation of different systems were performed to analyse the effect of warm starting on the convergence of NDEM simulations.  The following systems were studied: a 1D column,
formation of a pile, dense flow in a rotating drum and a triaxial shear cell.  
The main material and simulation parameters are listed in Table \ref{tab:parameters}.  The method was implemented in the software AgX 
Dynamics \cite{agx} in the module for NDEM simulation with optimized data
structures and support for collision detection and PGS using parallel 
computing on multicore processors.  The simulations were run on a 
desktop computer with Intel(R) Core(TM) i7 CPU, 2.8 GHz, 8 GB
RAM on a Windows 64 bit system.  Videos from simulations are found at 
\url{http://umit.cs.umu.se/granular/warmstarting/}.

\begin{table}
\caption{Main material and simulation parameters}
\label{tab:parameters}     
\begin{tabular}{lll}
\hline\noalign{\smallskip}
Notation & Value & Comment \\
\noalign{\smallskip}\hline\noalign{\smallskip}
$[d,d_2]$ 		& 	$[13, 10]$ m 		& bi-disperse particle diameter \\
$\rho$ 		& 	$3700$ kg/m$^3$ & particle mass density \\
$E$ 		& 	$6$ MPa \footnote{In the triaxial test $E = 60$ MPa$^3$ was used.} 	& Young's modulus \\
$e$ 		& 	$0.18$  		& restitution coefficient \\
$\textsub {\mu}{s}$ & $0.91$	& surface friction coefficient\\
$\textsub {\mu}{r}$ & $0.32$	& rolling resistance coefficient\\
$\Delta t$ 	& 	$5$ ms 			& 	timestep \\
$\textsub{v}{imp}$ 	& 	$0.05$ m/s & impact threshold \\
\noalign{\smallskip}\hline
\end{tabular}
\end{table}

\subsection{Column}
\label{sec:column}
\begin{figure}
  \includegraphics[width=0.45\textwidth]{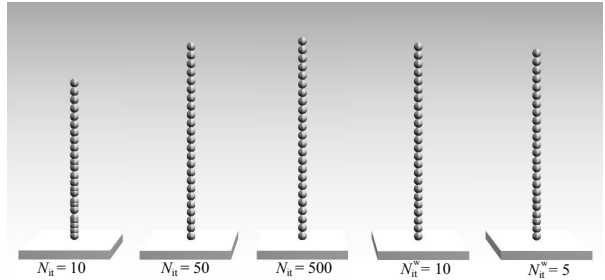}
\caption{Samples of five columns simulated, from left to right, with cold starting $\textsub{N}{it} = 10, 50$ and $500$, and history based warm starting $N^\text{w}_\text{it} = 10$ and $5$.}
 \label{fig:column_samples}
\end{figure}

\begin{figure}
  \includegraphics[width=0.45\textwidth]{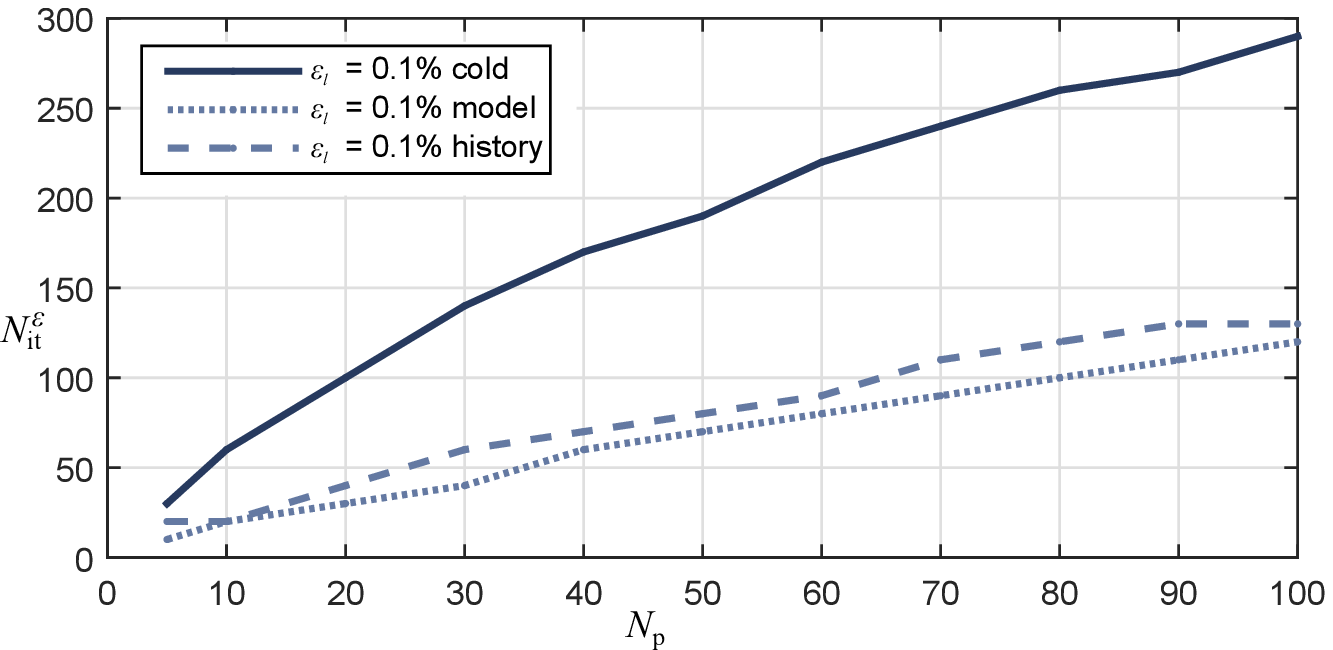}\\
  \includegraphics[width=0.45\textwidth]{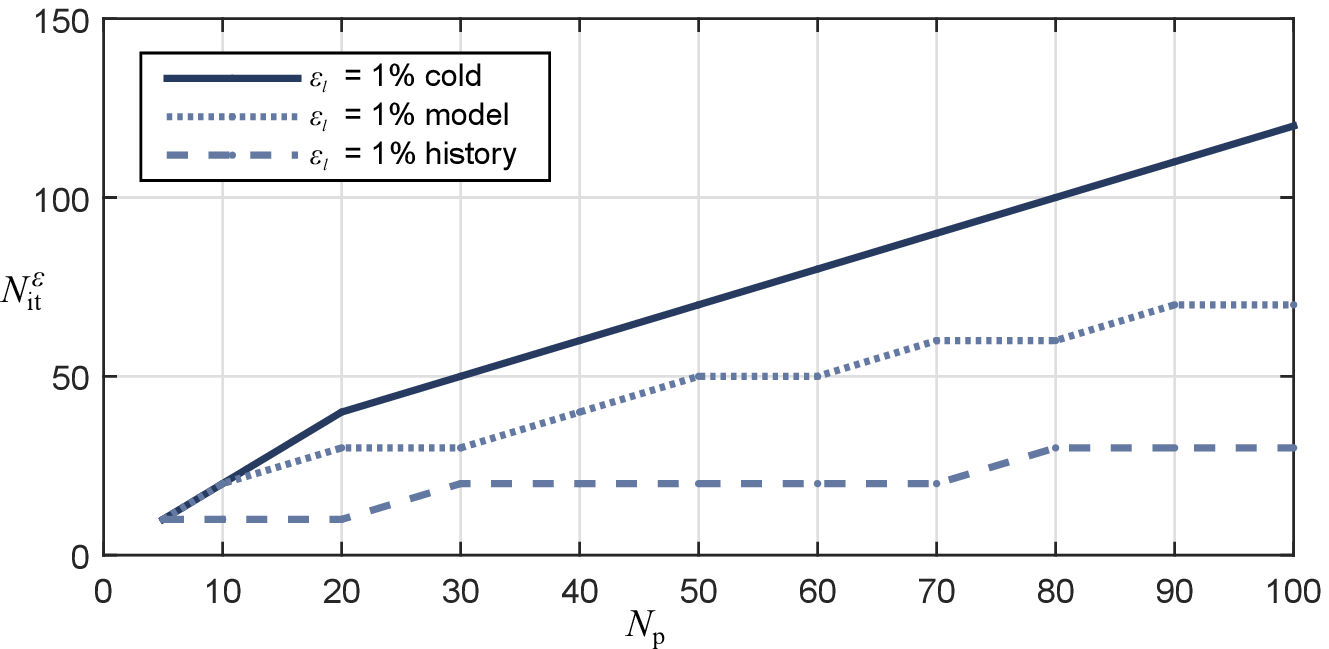}\\
  \includegraphics[width=0.45\textwidth]{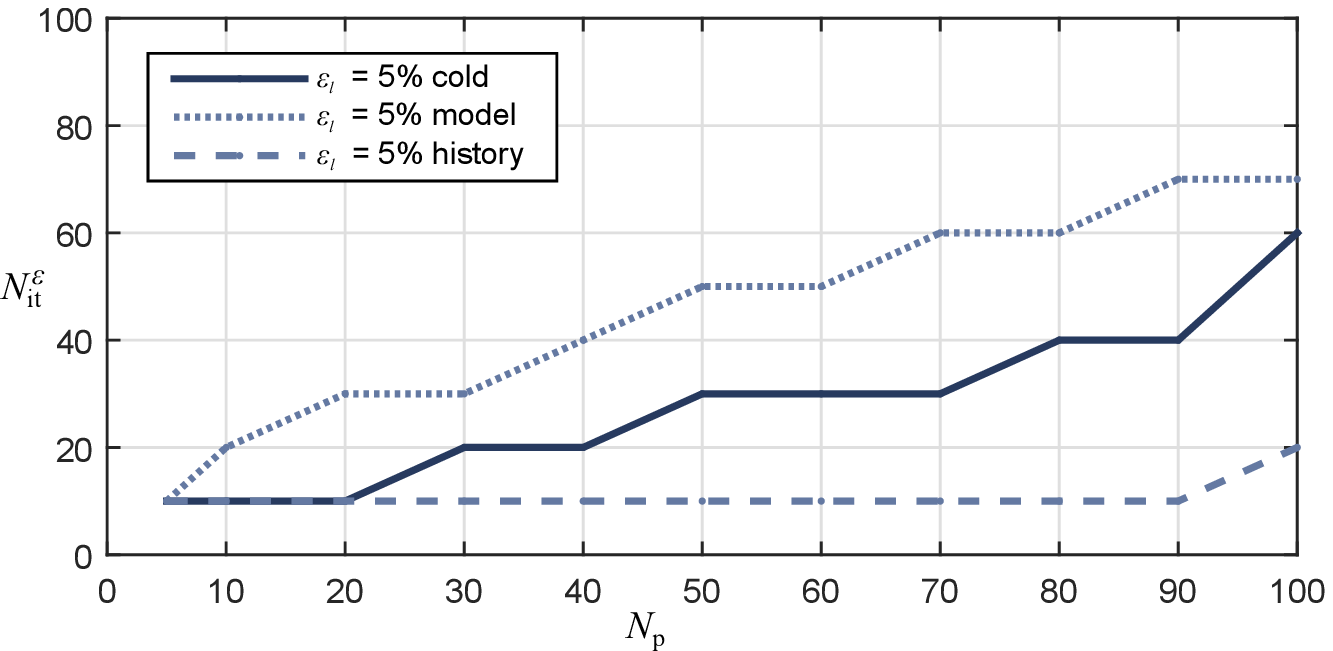}
\caption{The required number of iterations for a 1D column simulation for
error tolerance $\textsub{\varepsilon}{1D} = 0.1\%$ (top), $1 \%$ (middle) and $5 \%$ (bottom)
depending on the number of particles and warm starting method.}
 \label{fig:column_error_np_nit}
\end{figure}

\begin{figure}[h]
  \includegraphics[width=0.45\textwidth]{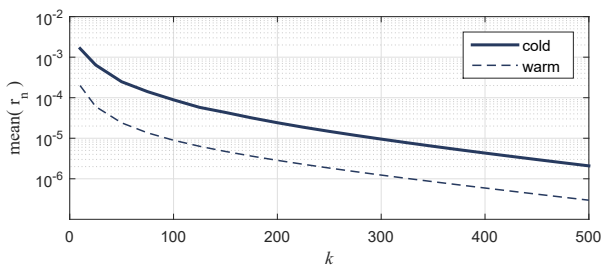}
\caption{The evolution of the mean normal force residual during a PGS solve for a $\textsub{N}{p} = 25$ column using cold starting and
history based warm starting.}
 \label{fig:column_residual}
\end{figure}

Particles of diameter $d = 13$ mm are initiated on top of another with zero 
overlap.  The system compress slightly under its weight.  The simulation is
run until the 1D column have come to rest.  Sample images
from simulation with and without warm starting and for
different number of iterations are shown in 
Fig.~\ref{fig:column_samples}.  Warm starting clearly improve
the convergence.
%
To make a quantitative convergence analysis we study the deviation of the
simulated column height, $l_{\textsub{N}{it}}$, from the theoretical 
height, $l$, computed using the Hertz contact law
\begin{equation}\label{eq:height}
	\textsub{\varepsilon}{l} = \frac{l - l_{\textsub{N}{it}}}{l}
\end{equation}
A series of simulations are run with number of particles, 
$\textsub{N}{P}$, ranging from $5$ and $100$, 
number of iterations, $\textsub{N}{it}$, ranging from $10$ to $500$.
The required number of iterations, $\textsub{N}{it}^{\varepsilon}$, to reach a solution with error 
tolerance $\textsub{\varepsilon}{l} = 0.1\%, 1\%$ and $5\%$
are presented in Fig.~\ref{fig:column_error_np_nit}.  It scales almost linearly with
the number of particles and increase with decreasing error tolerance $\textsub{\varepsilon}{l}$.
History based warm starting is on average three times as efficient as cold starting.  Also model based warm starting improve the convergence at low error tolerance.  The performance gain from model based warm starting decrease
with increasing error tolerance and for $\textsub{\varepsilon}{l} = 5\%$
model based warm starting require twice as many iterations as cold starting.
Figure \ref{fig:column_residual} show the evolution of the mean residual, see Eq.~(\ref{eq:residual}), for
the normal force constraint during a PGS solve for a column with $\textsub{N}{p} = 25$.  The 
convergence rates are similar but warm starting clearly has the advantage of starting closer to the solution.

\subsection{Pile formation}
\label{sec:pile}
\begin{figure}[h]
  \includegraphics[width=0.45\textwidth]{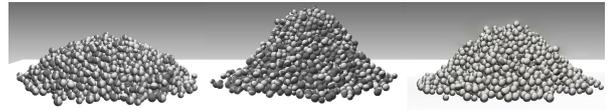}
\caption{Samples from simulations of pile formation.  From left
to right is the cold started pile ($\Delta t = 5$ ms, $\textsub{N}{it} = 50$), a reference pile ($\Delta t = 0.2 $ ms, $\textsub{N}{it} = 500$)
and a warm started pile ($\Delta t = 5$ ms, $\textsub{N}{it} = 50$)}
 \label{fig:pile_samples}
\end{figure}

\begin{figure}[h]
  \includegraphics[width=0.45\textwidth]{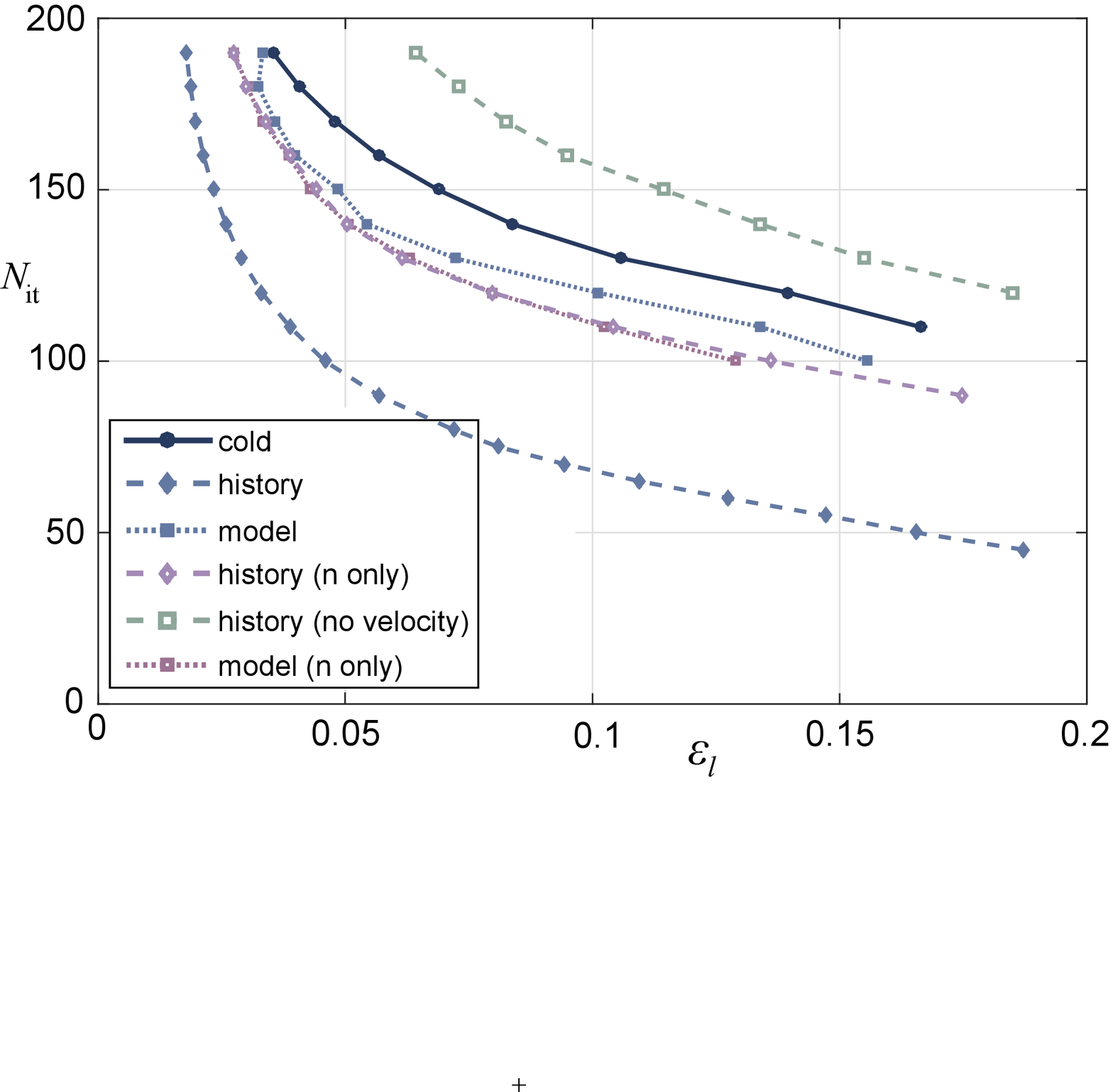}
\caption{The required number of iterations versus pile height error for different warm starting methods.}
 \label{fig:pile_height_error_it}
\end{figure}

\begin{figure}[h]
  \includegraphics[width=0.45\textwidth]{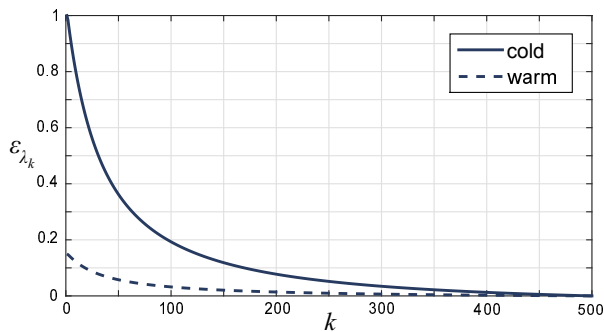}
\caption{The evolution of the relative multiplier during a PGS solve for a resting pile using cold starting and
history based warm starting.}
 \label{fig:pile_multiplier}
\end{figure}

\begin{figure}[h]
  \includegraphics[width=0.45\textwidth]{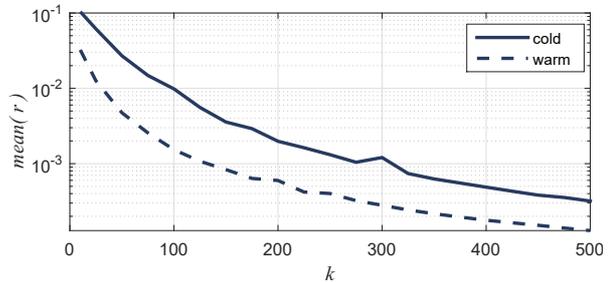}
\caption{The mean residual dependency on the number of iterations when
simulating a resting pile for $1$ s using cold starting and
history based warm starting.  }
 \label{fig:pile_residual}
\end{figure}

\begin{figure}[h]
  \includegraphics[width=0.45\textwidth]{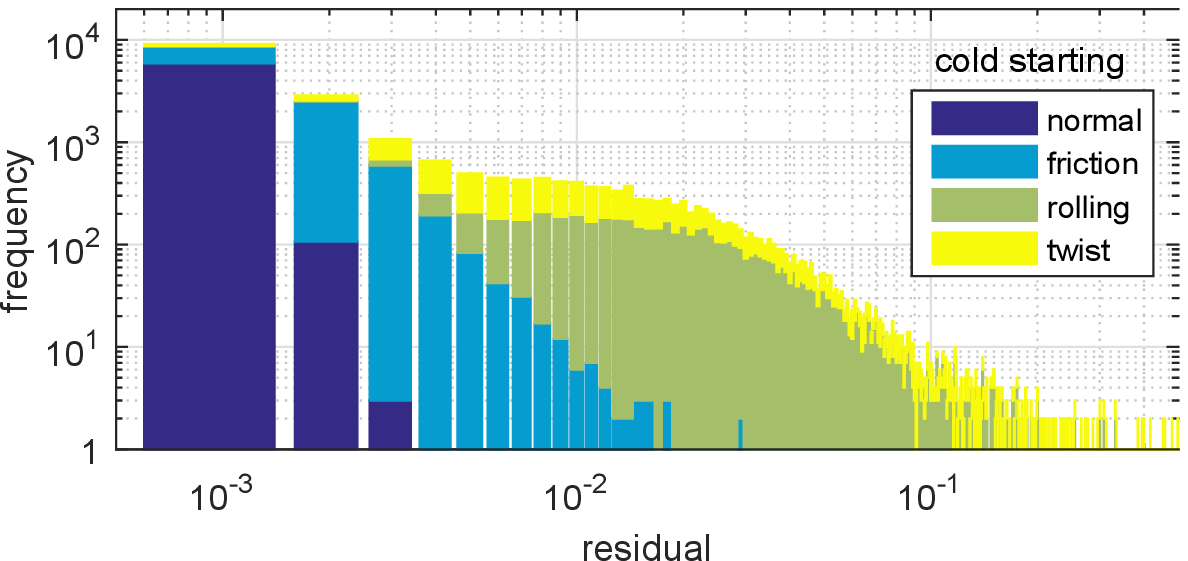}\\
    \includegraphics[width=0.45\textwidth]{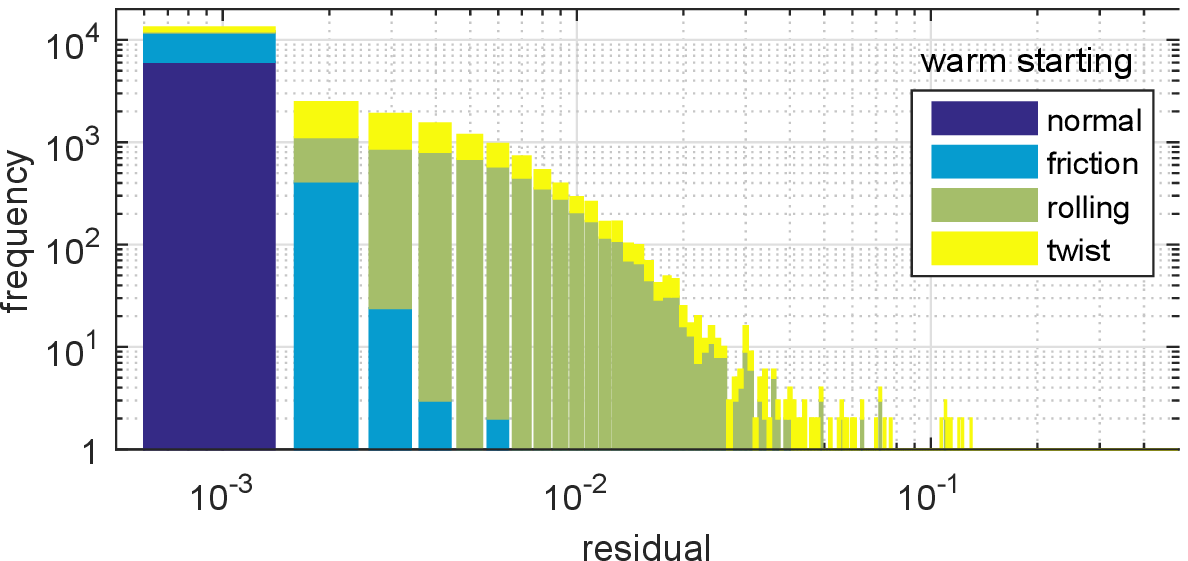}
\caption{The residual distribution for a resting pile after $1$ s using $\textsub{N}{it} = 100$ iterations,
cold starting (top) and history based warm starting (bottom).  }
 \label{fig:pile_residual_hist}
\end{figure}

A pile is formed by continuously emitting particles of diameter $d = 13$ mm from a $3d$ wide source placed
$20 d$ above a ground plane.  The number of particles in the pile is $\textsub{N}{p} = 3363$.    
Again we use the relative height, $\textsub{\varepsilon}{l}$ in Eq.~(\ref{eq:height}), as error measurement.  
The reference height about $15 d$ is measured from a pile constructed using small time-step $\Delta t = 0.2$ ms and 
$\textsub{N}{it} = 500$.  Pile formation is then simulated using time-step $\Delta t = 5$ ms for
different number of iterations and warm starting methods.  Sample images
from simulations are shown in Fig.~\ref{fig:pile_samples}.  
The angle of repose is an alternative measure but was found to give less precise result.
The pile height is measured $10$ s after the last emitted particle has come to approximate rest.  
Simulations are run with warm starting applied both to normal forces, friction and rolling resistance and to 
normal forces only. The historical warm starting is tested with and without the velocity
update associated with the warm start in Eq.~(\ref{eq:history_warmstart}).
The required number of iterations for a given error threshold are given in
Fig.~\ref{fig:pile_height_error_it}.  With few iterations the piles experience
artificial compression and contact sliding such that the pile gradually melt down to a singe particle layer.  The pile stability increase with the 
number of iterations.  History based warm starting, applied to both normals, friction and rolling resistance, give the best result  and require roughly 
half the number of iterations of cold starting.  If the warm start 
velocity is not applied the result is worse than cold starting.
Model based warm starting is only marginally better than cold starting and
is from further experiments here on excluded.
Applying warm starting to the normal constraints only does not improve
the convergence significantly.  
The convergence is also analysed by studying the evolution of the Lagrange
multiplier and the residual.  The relative error of the normal force 
multiplier is computed as
\begin{equation}
	\varepsilon_{\lambda_k} = \left\langle \frac{|\bg{\lambda}^{\text{n}(\alpha)}_{500} - \bg{\lambda}^{\text{n}(\alpha)}_k |}{|\bg{\lambda}^{\text{n}(\alpha)}_{500}|} \right\rangle
\end{equation}
The evolution of $\varepsilon_{\lambda_k}$ during a solve of a stationary pile
is shown in Fig.~\ref{fig:pile_multiplier}.  The multiplier
error for history based warm starting is roughly five times smaller than for cold starting and remain more accurate indefinitely.  
A more careful analysis can be made by studying the evolution of the residual, defined in 
Eq.~(\ref{eq:residual}), and how it is distributed over the
constraints.  To get comparable states a stationary pile is prepared
by using 500 iterations from which the cold and warm started 
simulations are started and run for $1$ s before the measurement.
The evolution of the mean residual during a PGS solve is shown in Fig.~\ref{fig:pile_residual}.  
The convergence rates are similar but the initial lead of history based warm starting over cold starting by roughly a factor $5$ remains throughout the 500 PGS iterations.  Comparing the residual histograms from using cold and warm starting
in Fig.~\ref{fig:pile_residual_hist}
it is clear that the solutions differs primarily in the errors for 
the rolling resistance and friction constraints and less so for
normal force constraints.  This is consistent with the faster melting
of the piles simulated with cold starting.

\subsection{Rotating drum}
\label{sec:drum}
\begin{figure}[h]
  \includegraphics[width=0.45\textwidth]{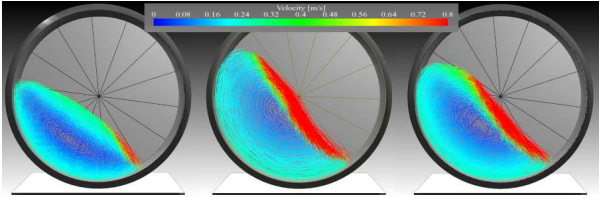}
\caption{A sample of particle trajectories from simulation of a rotating drum with $\Omega = 0.25$ rad/s, $\Delta t = 5$ ms and $\textsub{N}{it} = 10$ (left), $\textsub{N}{it} = 500$ (middle) and warm starting $\textsub{N}{it} = 10$ (right).}
 \label{fig:drum_trajectories}
\end{figure}

\begin{figure}[h]
  \includegraphics[trim=0cm 0cm 7mm 0cm, clip=true, width=0.45\textwidth]{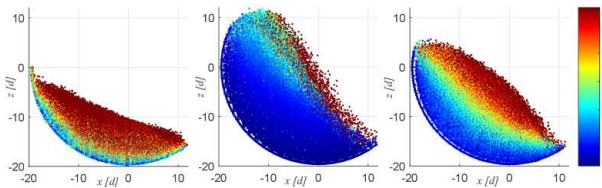}
\caption{A sample of cross-section flow from a simulation of a rotating drum with $\Omega = 0.25$ rad/s, $\Delta t = 5$ ms and $\textsub{N}{it} = 10$ (left), $\textsub{N}{it} = 500$ (middle) and warm starting $\textsub{N}{it} = 10$ (right).  The colour coding show the particle velocities 
relative to rigid co-motion with the drum.}
 \label{fig:drum_cross}
\end{figure}

\begin{figure}[h]
  \includegraphics[width=0.45\textwidth]{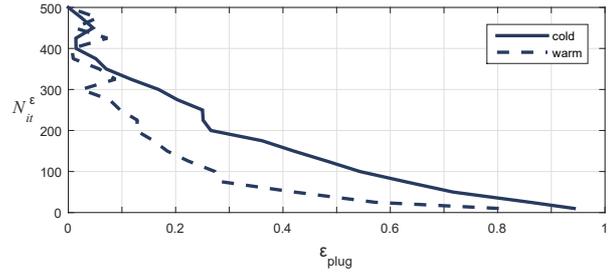}
\caption{The convergence of the plug zone fraction number for
cold starting and history based warm starting.}
 \label{fig:drum_plug_convergence}
\end{figure}

\begin{figure}[h]
  \includegraphics[width=0.45\textwidth]{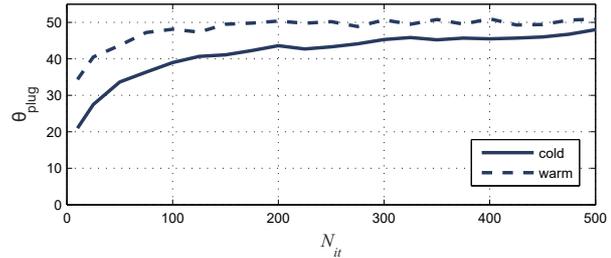}
\caption{The dynamic angle of repose as function of
number of iterations for cold starting and history based warm starting.}
 \label{fig:drum_repose}
\end{figure}

A cylindrical drum with diameter $D = 40 d$ and width $w = 7d $ is rotated with angular velocity $\Omega = 0.25$ rad/s. 
This corresponds to the Froude number $\text{Fr} \equiv D\Omega^2/2g \sim10^{-3}$ which
corresponds to the dense rolling flow regime.  A nearly stationary flow of $\textsub{N}{p} = 4864$ particles with bi-disperse
size distribution $d$ and $d_2$.  At this low Froude number a large plug-zone is developed where particles co-rotate
rigidly with the drum.   A convergence analysis is made of the plug-zone number fraction, $\textsub{N}{plug}/\textsub{N}{p}$, 
and the dynamic angle of repose,
$\theta'$.  These are measured for different number of iterations on a flow averaged over $2$ s for cold starting and 
historical warm starting.
The sample trajectories in Fig.~\ref{fig:drum_trajectories} illustrate the general trend that the dynamic angle of
repose and the size of the plug zone decrease with decreasing number of iterations but less so using warm starting.
The normalized particle flow velocity relative the plug flow is computed
$\textsub{v}{r}^i \equiv |\tv{v}^i - \tv{r}^i \times \bg{\Omega}|/R\Omega$ 
and sample plots are shown in Fig.~\ref{fig:drum_cross}.  
As threshold for the plug zone flow we set $\textsub{v}{r} \leq 0.15$, which is fulfilled by
$\textsub{N}{plug}^{500}/\textsub{N}{p} = 58\%\pm5\%$ particles where the
variations reflect the slightly pulsating nature of the flow, 
due to sequential onset of avalanches.  The plug zone fraction number error is defined
\begin{equation}
	\textsub{\varepsilon}{plug} = \frac{\textsub{N}{plug}^{500} - N_\text{plug}^{\textsub{N}{it}}}{\textsub{N}{p}}
\end{equation}
and the relation to the required number of iterations is found in
Fig.~\ref{fig:drum_plug_convergence}.  The warm starting solution 
approach the solution faster but seems to have larger variations 
at high iteration numbers.  
The convergence analysis of the dynamic angle of repose also
show that warm starting converges faster although to a slightly higher angle $\textsub{\theta}{w,plug}^{500} = 50^{\circ}$
compared to  $\textsub{\theta}{plug}^{500} = 48^{\circ}$, see Fig.~\ref{fig:drum_repose}.  The dynamic angle of repose is measured 
as the displacement of the material centre of mass from the $z$-axis
which is more robust than tracking the surface. 

\subsection{Triaxial shear}
\label{sec:collapse}
\begin{figure}[h]
  \includegraphics[width=0.35\textwidth]{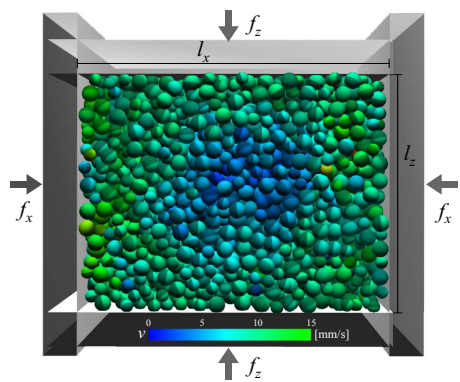}
\caption{Sample image from the triaxial test.}
 \label{fig:triaxial}
\end{figure}

\begin{figure}[h]
  \includegraphics[width=0.45\textwidth]{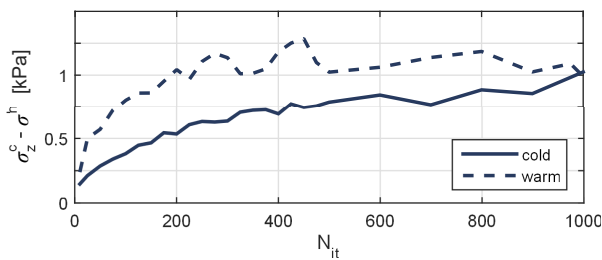}
\caption{Critical yield stress as function of number of iterations for cold starting and
history based warm starting.}
 \label{fig:mean_stress_iteration}
\end{figure}

\begin{figure}[h]
  \includegraphics[width=0.45\textwidth]{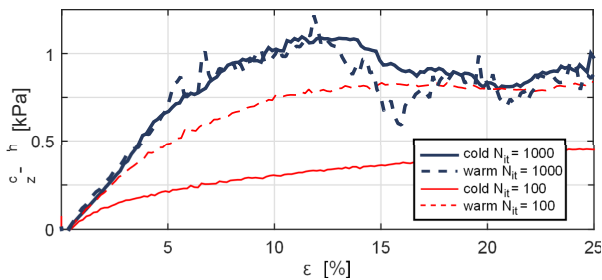}
\caption{Sample stress curves in triaxial test for 100 and 1000 iterations. }
 \label{fig:stress_curves}
\end{figure}

The triaxial shear test is constructed by six dynamic rigid walls of
mass $100$ kg each that are driven with prismatic motors to apply
a specific stress $\sigma_i = f_i / A_i$, where $A_i$ is the
cross-section area and $f_i$ the applied motor force in the
coordinate direction $i = x, y, z$, see Fig.~\ref{fig:triaxial}  
First, a hydrostatic pressure
of $\textsup{\sigma}{h} = 100$ Pa is applied on all sides. Then the top and bottom
walls are driven inwards at $0.01$ m/s by regulating $\sigma_z$ and
maintaining a constant side wall pressure at $\sigma_x = \sigma_y = \textsup{\sigma}{h}$.  At some
critical deviator stress $\textsup{\sigma}{c}_z - \textsup{\sigma}{h}$ the material fail
to sustain further increase in stress and starts to  shear indefinitely.  The transition is
more or less sharp depending on the initial packing ratio, hydrostatic pressure and applied shear rate.
In this test the Young's modulus is set to the stiffer value of $E = 60$ MPa to get a 
sharper transition between compression and shear.
The critical axial stress $\textsup{\sigma}{c}_z$ is computed as the averaged
$\sigma_z$ in the shear phase between lateral strain $\varepsilon = 10\% $
to $\varepsilon = 25\% $.  
The critical stress deviator depending on the number of iterations for
cold starting and history based warm starting is shown in Fig.~\ref{fig:mean_stress_iteration}.
Both curves converge to about $1 \pm 0.2$ kPa.  With warm starting
the stress levels out at $\textsub{N}{it} \gtrsim 200$ while cold starting
require $\textsub{N}{it} = 1000$.  
Sample curves of the stress deviator as function of lateral strain
are shown in Fig.~\ref{fig:stress_curves}.  These confirm the faster convergence
when warm starting but also show higher stress fluctuations in the shear phase.
Whether this is an artefact of the warm starting or an actual feature of the triaxial
test has not been pursued.

\section{Application example}
\label{sec:application}
\begin{figure}[h]
  \includegraphics[width=0.45\textwidth]{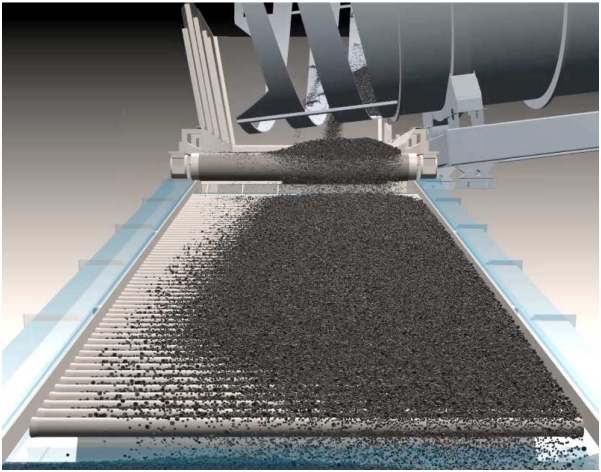}
\caption{A balling drum circuit with granular material in different states.}
 \label{fig:balling_circuit}
\end{figure}

The effect of using warm starting in practical simulation applications is illustrated with two
examples.  The first
example is part of a balling drum circuit used in ore pelletizing systems \cite{wang:2015:pvn}, see Fig.~\ref{fig:balling_circuit}.
Simulations are used for the purpose of process control and for finding a design of the 
drum outlet that maximizes an evenly distributed throughput
on a roller sieve where material is size distributed.
Three distinctive subsystems with different dynamics can be 
identified.  Firstly, there is the drum with an almost stationary flow.  Secondly, material is distributed onto quasistationary 
piles on a wide-belt conveyor.  Thirdly, the particles disperse over a roller sieve with increasing gap size downwards
to achieve a size separation.  The design problem is fore mostly a geometric flow problem and the material 
distribution need to be computed with sufficient accuracy.  We assume $5$ \% is a required accuracy
for dynamic and static angle of repose.  From Fig.~\ref{fig:drum_repose} we estimate that warm starting
is roughly three times more computationally efficient in computing the drum flow and, according to
Fig.~\ref{fig:pile_height_error_it}, twice as efficient for pile formation on the conveyor.  The flow
on the roller sieve is more disperse and collision dominated requiring only few iterations ($\textsub{N}{p} < 25$)
and it can be expected that warm and cold starting are equally efficient.  The overall computational
speed-up by applying warm starting is thus estimated to a factor no larger than 2.

\begin{figure}[h]
  \includegraphics[width=0.45\textwidth]{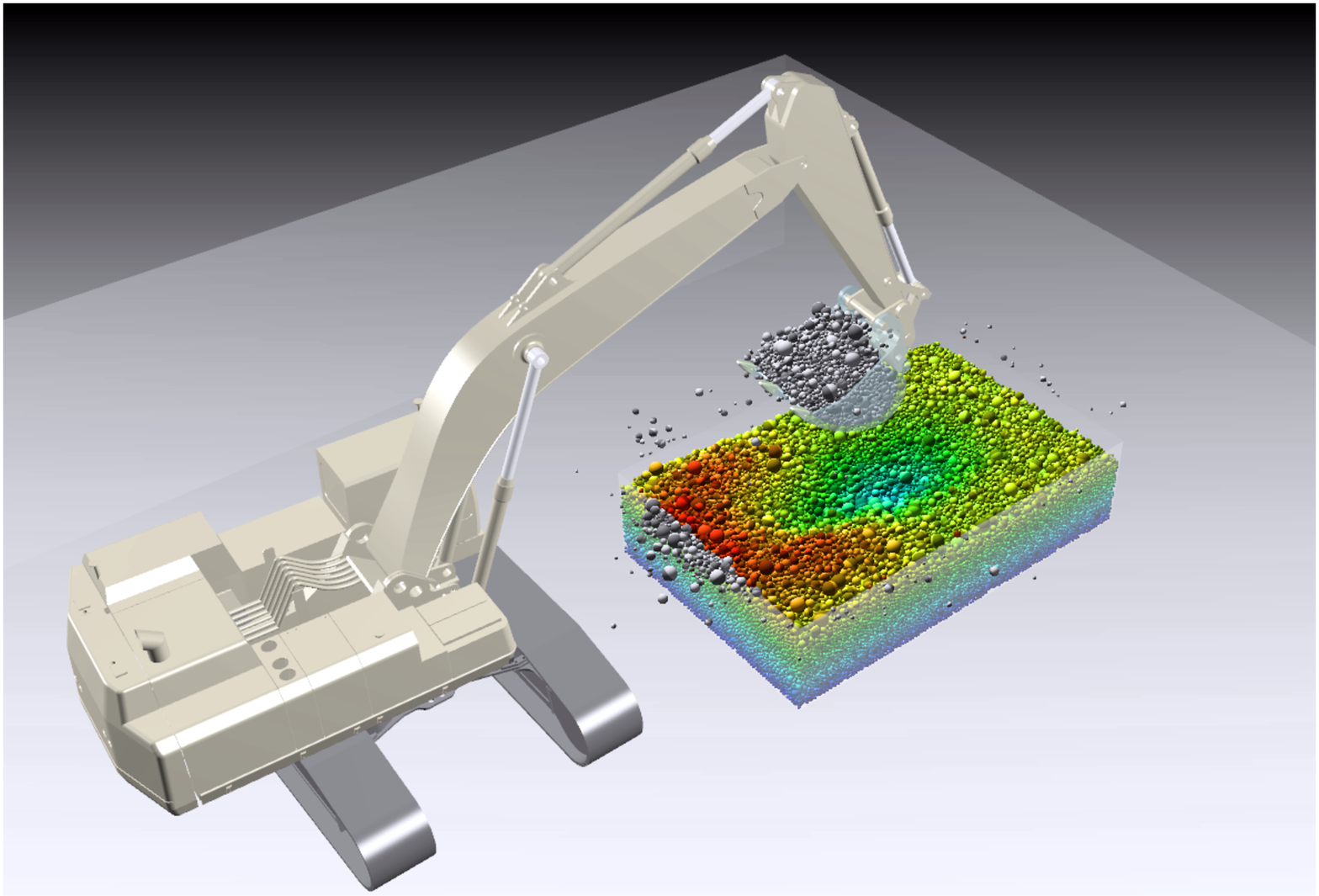}\\ \hspace{3 mm}
  \includegraphics[width=0.45\textwidth]{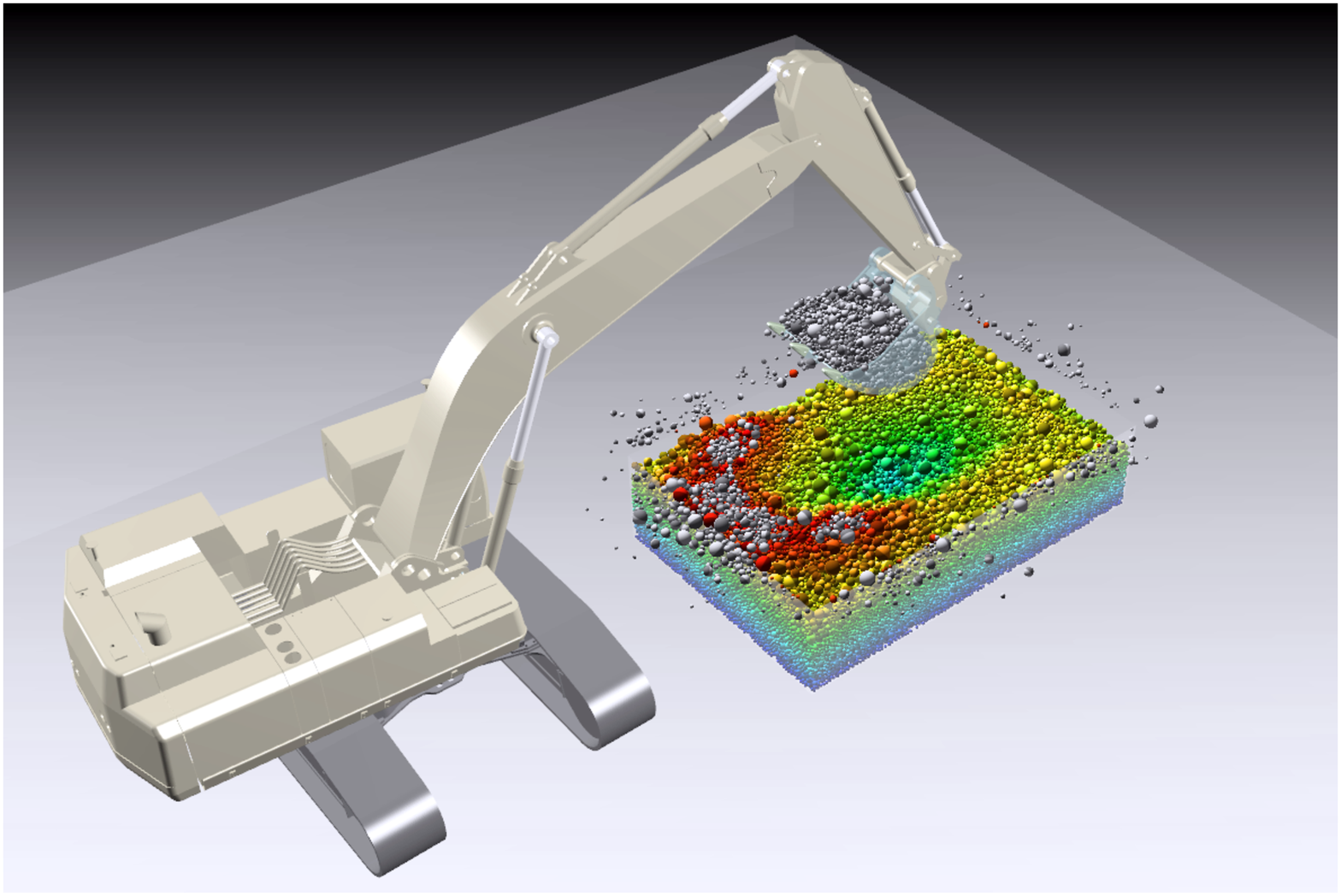}
\caption{An excavator digging in trench with $~10^5$ particles, $h = 2.5 ms$, $\textsub{N}{it} = 25$ and
using cold starting (top) and warm starting (bottom).  The colour codes the particle height with red to blue
ranging from $0$ m to $-2$ m.  Gray particles are above $0$ m.}
 \label{fig:excavator}
\end{figure}

\begin{figure}[h]
  \includegraphics[width=0.45\textwidth]{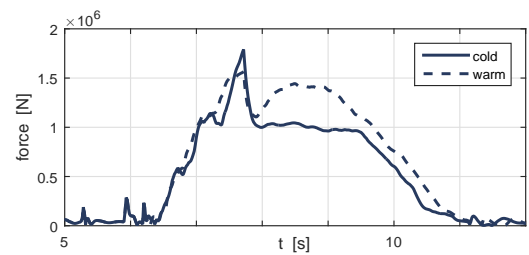}
\caption{The force trajectory of the middle link pistons of the excavator while digging with
$\textsub{N}{it} = 25$ using cold starting and warm starting.}
 \label{fig:link_force}
\end{figure}

The second example is an excavator.  A rectangular trench is filled with roughly $10^5$
spherical particles of uniform size distribution between $25$ and $100$ mm
and particle mass density of $2500$ kg/m$^3$.  The excavator is modeled as a rigid 
multibody system of total mass $50$ ton divided in 10 bodies, 8 joints and 3 linear  
actuators (hydraulic cylinders) and one rotational motor.  
The full system of granulars and vehicle take the mathematical form of
Eq.~(\ref{eq:MLCP}) and is solved using a split solver where the vehicle part is solved
using a direct block-sparse pivoting method \cite{agx} and the granular material with a
PGS solver as described in this paper.  Simulations were run with time-step
$h = 2.5$ ms, which allow for a low number of iterations.  The machine perform an
excavation cycle by a pre-programmed control signal to the actuators.
The resulting actuator forces are measured and these include the back 
reaction from the resistance and inertia of the granular material.  
Two simulations, with and without warm starting,
are run with $\textsub{N}{it} = 25$.  Sample images from the simulations are shown
in Fig.~\ref{fig:excavator}.  Observe the difference in height surface of the granular material due to artificial compression and frictional slippage due
to numerical errors in the PGS solve.  The undisturbed height in the two
simulations differ 
by $10$ \% and volume of displaced material differ by at least $30$ \%.
The difference in granular dynamics also affect the measured force response.
The force trajectory of the middle actuator is provided in Fig.~\ref{fig:link_force}.
In the phase between $8\--10$ s, when the bucket is dragged through the 
material the force when using warm starting is almost $50\%$ larger because
more material is set in motion and stronger resistance to shear motion.
Whether $h = 2.5$ ms, $\textsub{N}{it} = 25$ and the improvement
by using warm starting give sufficiently accurate force response depend on the intended use of the data and 
require further convergence analysis.  On a desktop
computer\footnote{Performance measurement are made on a desktop computer with
Intel(R) Core(TM) Xeon X5690, 3.46 GHz, 48 GB RAM on a Linux 64 bit system.} 
with the given NDEM settings the computational time is roughly $100$ s per
realtime second.

\section{Conclusions}
\label{sec:conclusions}

The convergence of the projected Gauss-Seidel algorithm
for NDEM simulation is increased by warm starting with the solution 
from previous time-step.  The computational speed-up by warm starting is
demonstrated to be about $2\--5$ for a wide range of systems including pile formation, 
granular drum flow and triaxial shear.  An examination of the residual distribution show that convergence improvement 
primarily improve on the velocity constraints \-- friction and rolling resistance \--
and less so on the normal force constraints.  Warm starting the Lagrange multiplier based on 
$85$ \% of the value from last  time-step was found to give best results.  
Warm starting based on an explicit contact force model give only marginal 
speed-up, for example $20$ \% for a pile formation.  This is not surprising since the 
damping coefficients in the dissipation models for sliding and rolling are not physics
based and can only predict the value of contact forces in slide mode but not of
stick mode inside the friction and rolling resistance limits.  For materials shearing
under high stress, compared
to the stress produced by the materials own weight, warm starting show larger
stress fluctuations than without warm starting.  Whether this is an artefact or correct behaviour
emerging after further iterations has not been established.  A more in depth analysis of systems under large stress should be made 
considering also alternative size of time step, shear rate, hydrostatic load stress and particle
stiffness.  

\begin{acknowledgements}
This project was supported by Algoryx Simulations, LKAB, UMIT Research Lab and
VINNOVA (dnr 2014-01901).
\end{acknowledgements}

\section*{Appendix}
\label{sec:appendix}

\subsection*{A. Simulation algorithm}
The algorithm for simulating a system of granular material using
NDEM with PGS solver with warm starting is given in Algorithm \ref{alg:NDEMPGS}.
\begin{algorithm}[h!]
  \caption{NDEM simulation with warm started PGS solver}\label{alg:NDEMPGS}
  \begin{algorithmic}[1]
  \State set constants and parameters
  \State initial state: $(\tv{x}_{0}, \tv{v}_{0})$
  \For{$i = 0,1,2,\hdots,t/\Delta t$} \Comment{Time stepping}
  \State contact detection
  \State compute $\tv{g}, \tv{G}, \bg{\Sigma}, \tv{D}$
  \State impact stage PGS solve $\tv{v}_{i} \to (\tv{v}^{+}_{i},\bg{\lambda}^{+}_{i})$   \Comment{impacts}
  \State compute $\textsub{\tv{q}}{n} = -(4/\Delta t) \textsub{\bg{\Upsilon}}{n}\textsub{\tv{g}}{n} + \textsub{\bg{\Upsilon}}{n}\textsub{\tv{G}}{n}\tv{v}^{+}_i$
  \State pre-step $\tv{v} = \tv{v}^{+}_{i} + \Delta t \tv{M}^{-1} \textsub{\tv{f}}{ext}$ 
  \State $\bg{\lambda}_{k_0} = \tv{0}$ or  warm start $\bg{\lambda}_{k_0}$ 
  \State warm-step $\tv{v} = \tv{v} + \tm{M}^{-1}\tm{G}^\T\bg{\lambda}_{k_0}$ 
  \State $---$ \Comment{PGS solve for continuous contacts$----$}
  \For{$k = 1,\hdots,\textsub{N}{it}$ and {\bf while} \emph{criteria}$(\tv{r})$} 
  \For{each contact $\alpha = 0,1,\hdots,\textsub{N}{c}-1$}
  \For{each constraint $n$ of contact $\alpha$}
	  \State $\tv{r}^{(\alpha)}_{n,k} = - \tv{q}^{(\alpha)}_{n,k} + \tv{G}^{(\alpha)}_{n}\tv{v}$ \Comment{residual}
	  \State $\bg{\lambda}^{(\alpha)}_{n,k} = \bg{\lambda}^{(\alpha)}_{n,k-1} + \tv{D}^{-1}_{n,(\alpha)} \tv{r}^{(\alpha)}_{n,k}$ \Comment{multiplier}
	  \State $\bg{\lambda}^{(\alpha)}_{n,k} \leftarrow \text{proj}_{\mathcal{C}_{\mu}}(\bg{\lambda}^{(\alpha)}_{k})$ \Comment{project}
	  \State $\Delta \bg{\lambda}^{(\alpha)}_{n,k} =  \bg{\lambda}^{(\alpha)}_{n,k} - \bg{\lambda}^{(\alpha)}_{n,k-1}$
	  \State $\tv{v}  =  \tv{v} + \tv{M}^{-1}\tv{G}^{T}_{n,(\alpha)}\Delta\bg{\lambda}^{(\alpha)}_{n,k}$
  \EndFor
  \EndFor
  \EndFor
  \State $\tv{v}_{i+1}  = \tv{v}$	\Comment{velocity update}
  \State $\tv{x}_{i+1}  = \tv{x}_i + \Delta t \tv{v}_{i+1}$	\Comment{position update}
  \EndFor
  \end{algorithmic}
\end{algorithm}
Based on the Hertz contact law, each contact $\alpha$ between body $a$ and $b$ add contributions to the constraint vector and
normal and friction Jacobians according to
\begin{eqnarray}
\label{eq:constraint}
\delta_{(\alpha)} & = & \mathbf{n}_{(\alpha)}^\T(\mathbf{x}_{a} + \mathbf{d}^{(\alpha)}_{a} - \mathbf{x}_{b} - \mathbf{d}^{(\alpha)}_{b})\nonumber\\
g_{(\alpha)} & = & \delta^{\textsub{e}{H}}_{(\alpha)}\quad,\ \textsub{e}{H} = 5/4 \nonumber\\
\tm{G}_{\text{n}a}^{(\alpha)} & = & \textsub{e}{H} g^{\textsub{e}{H}-1}_{(\alpha) } \left[ 
	\begin{array}
	      		[c]{cc}%
	      		- \mathbf{n}_{(\alpha)}^{\T} &\ \  -(\mathbf{d}^{(\alpha)}_{a} \times \mathbf{n}_{(\alpha)})\tran{T}	    	
	\end{array} \right]\nonumber\\
\tm{G}_{\text{n}b}^{(\alpha)} & = & \textsub{e}{H} g^{\textsub{e}{H}-1}_{(\alpha) } \left[ 
	\begin{array}
	      		[c]{cc}%
	      		\mathbf{n}_{(\alpha)}^{\T} &\ \  (\mathbf{d}^{(\alpha)}_{b} \times \mathbf{n}_{(\alpha)})\tran{T}
	\end{array} \right]\\
\tm{G}_{\text{t}a}^{(\alpha)} & = & \left[ 
	\begin{array}
	      		[c]{cc}%
	      		- \mathbf{t}^{(\alpha)\T}_1 &\ \  -(\mathbf{d}^{(\alpha)}_{a} \times \mathbf{t}^{(\alpha)}_1)\tran{T}	    	\\
	      		- \mathbf{t}^{(\alpha)\T}_2 &\ \  -(\mathbf{d}^{(\alpha)}_{a} \times \mathbf{t}^{(\alpha)}_2)\tran{T}
	      		\end{array} \right]\nonumber\\
\tm{G}_{\text{t}b}^{(\alpha)} & = & \left[ 
	\begin{array}
	      		[c]{cc}%
	      		\mathbf{t}^{(\alpha)\T}_1 &\ \  (\mathbf{d}^{(\alpha)}_{b} \times \mathbf{t}^{(\alpha)}_1)\tran{T}	    	\\
	      	 	\mathbf{t}^{(\alpha)\T}_2 &\ \  (\mathbf{d}^{(\alpha)}_{b} \times \mathbf{t}^{(\alpha)}_2)\tran{T}
	      		\end{array} \right]\nonumber\\
	\label{eq:Jacobian_rolling}
	\tm{G}_{\text{r}a}^{(\alpha)} & = &
	\left[
	\begin{array}
	      		[c]{cccc}%
				\mathbf{0}_{1\times 3} &\ \ \mathbf{t}^{(\alpha)\T}_1 &\ \  \mathbf{0}_{1\times 3} & -\mathbf{t}^{(\alpha)\T}_1\\
				\mathbf{0}_{1\times 3} &\ \ \mathbf{t}^{(\alpha)\T}_2 &\ \  \mathbf{0}_{1\times 3} & -\mathbf{t}^{(\alpha)\T}_2\\
				\mathbf{0}_{1\times 3} &\ \ \mathbf{n}^{(\alpha)\T} &\ \  \mathbf{0}_{1\times 3} & -\mathbf{n}^{(\alpha)\T}\\
	\end{array} \right]\nonumber\\
	\tm{G}_{\text{r}b}^{(\alpha)} & = &
	\left[
	\begin{array}
	      		[c]{cccc}%
				\mathbf{0}_{1\times 3} &\ \ -\mathbf{t}^{(\alpha)\T}_1 &\ \  \mathbf{0}_{1\times 3} & \mathbf{t}^{(\alpha)\T}_1\\
				\mathbf{0}_{1\times 3} &\ \ -\mathbf{t}^{(\alpha)\T}_2 &\ \  \mathbf{0}_{1\times 3} & \mathbf{t}^{(\alpha)\T}_2\\
				\mathbf{0}_{1\times 3} &\ \ -\mathbf{n}^{(\alpha)\T} &\ \  \mathbf{0}_{1\times 3} & \mathbf{n}^{(\alpha)\T}\\
	\end{array} \right]\nonumber
\end{eqnarray}
where $\tv{d}^{(\alpha)}_{a}$ and $\tv{d}^{(\alpha)}_{b}$ are the positions of the contact point $\alpha$ relative to the particle positions $\tv{x}_{a}$ and $\tv{x}_{b}$.  The orthonormal contact normal and tangent vectors are $\mathbf{n}^{(\alpha)}$, $\mathbf{t}^{(\alpha)_1}$ and $\mathbf{t}^{(\alpha)_2}$.   

The diagonal matrices and Schur complement matrix $\tv{D}$ are
\begin{eqnarray}
	  \label{eq:diagonal}
	\textsub{\bg{\Sigma}}{n} & = & \frac{4}{\Delta t^2}\frac{\textsub{\varepsilon}{n}}{1+4\tfrac{\textsub{\tau}{n}}{\Delta t}}\bm{1}_{N_{c}\times N_{c}}\nonumber\\ 
	\textsub{\bg{\Sigma}}{t} & = & \frac{\textsub{\gamma}{t}}{\Delta t}\bm{1}_{2N_{c}\times 2N_{c}}\nonumber\\
	\textsub{\bg{\Sigma}}{r} & = & \frac{\textsub{\gamma}{r}}{\Delta t}\bm{1}_{3N_{c}\times 3N_{c}}\\
	\textsub{\bg{\Upsilon}}{n} & = & \frac{1}{1+4\tfrac{\textsub{\tau}{n}}{\Delta t}}\bm{1}_{N_{c}\times N_{c}}\nonumber\\
	\tv{D} & = & \tv{G}\tv{M}^{-1}\tv{G}^T + \bm{\Sigma}\nonumber
\end{eqnarray}
The mapping between MCP parameters and material parameters are
\begin{eqnarray}
	  \label{eq:regularization}
	\textsub{\varepsilon}{n} & = & \textsub{e}{\tiny H}/\textsub{k}{n} = 3\textsub{e}{\tiny H}(1-\nu^2)/E\sqrt{r^*} \nonumber\\ 
	\textsub{\tau}{n} & =&  \max(\textsub{n}{s}\Delta t, \textsub{\varepsilon}{n}/\textsub{\gamma}{n}) \\
	\textsub{\gamma}{n}^{-1} & = & \textsub{k}{n}c/e^2_{\text{\tiny H}}\nonumber
\end{eqnarray}
where $r^{*} = (r_a + r_b)/r_a r_b$ is the effective radius and we use $\textsub{\gamma}{t} = \textsub{\gamma}{r} = 10^{-6}$, $\textsub{n}{s} = 2$.



\end{document}